\documentclass[preprint,amsmath,amssymb,superscriptaddress,nobibnotes,bibnotes]{revtex4-1}

\usepackage{graphicx}
\usepackage{amssymb}
\usepackage{amsmath}
\usepackage{graphicx}
\usepackage{epsfig}
\usepackage{color}
\begin{document}

\title{An electronic origin of charge order in infinite-layer nickelates}
\author{Hanghui Chen}
\email{hanghui.chen@nyu.edu}
\affiliation{NYU-ECNU Institute of Physics, NYU Shanghai, Shanghai 200122, China}
\affiliation{Department of Physics, New York University, New York, New York 10012, USA}
\author{Yi-feng Yang}
\email{yifeng@iphy.ac.cn}
\affiliation{Beijing National Laboratory for Condensed Matter Physics and Institute of
Physics, Chinese Academy of Sciences, Beijing 100190, China}
\affiliation{University of Chinese Academy of Sciences, Beijing 100190, China}
\affiliation{Songshan Lake Materials Laboratory, Dongguan, Guangdong 523808, China}
\author{Guang-Ming Zhang}
\email{gmzhang@tsinghua.edu.cn}
\affiliation{State Key Laboratory of Low-Dimensional Quantum Physics and Department of
Physics, Tsinghua University, Beijing 100084, China}
\affiliation{Frontier Science Center for Quantum Information, Beijing 100084, China}
\author{Hongquan Liu}
\affiliation{NYU-ECNU Institute of Physics, NYU Shanghai, Shanghai 200122, China}
\date{\today}

\begin{abstract}

A charge order (CO) with a wavevector
$\mathbf{q}\simeq\left(\frac{1}{3},0,0\right)$ is observed in
infinite-layer nickelates. Here we use first-principles calculations
to demonstrate a charge-transfer-driven CO mechanism in infinite-layer
nickelates, which leads to a characteristic Ni$^{1+}$-Ni$^{2+}$-Ni$^{1+}$ stripe
state. For every three Ni atoms, due to the presence of
near-Fermi-level conduction bands, Hubbard interaction on Ni-$d$
orbitals transfers electrons on one Ni atom to conduction bands and
leaves electrons on the other two Ni atoms to become more
localized. We further derive a low-energy effective model to elucidate
that the CO state arises from a delicate competition between Hubbard
interaction on Ni-$d$ orbitals and charge transfer energy between Ni-$d$
orbitals and conduction bands. With physically reasonable parameters,
$\mathbf{q}=\left(\frac{1}{3},0,0\right)$ CO state is more stable than uniform paramagnetic state and usual
checkerboard antiferromagnetic state. Our work highlights the
multi-band nature of infinite-layer nickelates, which leads to some
distinctive correlated properties that are not found in cuprates.
\end{abstract}

\maketitle


\section*{Introduction}
Motivated by cuprates, superconductivity was proposed
~\cite{Anisimov-Rice-prb1999,Lee-Pickett-prb2004} and has been recently
discovered in hole-doped infinite-layer nickelates that contains
Ni$^{1+}$ ions of the similar $d^9$ electronic
configuration~\cite{DanfengLi-Nature2019,DanfengLi-prl2020,Zeng-prl2020,Osada-prm2020,hhwen2020,Osada2021,Zeng-SciAdv2022}.
While this seems to confirm the initial expectation and support the
Mott scenario for high-temperature superconductivity
~\cite{Lee-Nagaosa-Wen-rmp2006}, the nickelates show very different
properties even in their undoped parent compounds. Instead of an
antiferromagnetic Mott insulator, the resistivity of the parent
compounds exhibits metallic behavior at high temperatures and an
upturn below around 70 K~\cite{DanfengLi-Nature2019, Osada2020,Ikeda-apl2016, Osada2021}. So far no static long-range magnetic
order has been observed down to the lowest measured
temperature~\cite{Hayward-SSS2003, Zhao-2021}, although spin
fluctuations have been reported in
experiment~\cite{Lu-Science2021}. These are all different from
cuprates and have stimulated heated debate concerning the nature of
the nickelate parent state, from which superconductivity is born upon
hole doping. Although several theoretical scenarios
~\cite{Sakakibara-prl2020,Nomura-prb2019,Botana-Norman-prx2020,Kitatani2020,Karp-2022} have been proposed to emphasize the key role of single-band
Hubbard interaction,
a growing consensus is that multi-band physics is indispensable in
nickelates~\cite{Hepting-NatureMater2020,ZhangYangZhang-prb2020,Yang-Zhang2022,Wang-Zhang-Yang-Zhang,Hanghui-2020,Lechermann2020, Kang-2021, Hu-2019, Werner-2020,Zhang-2020,Wan-2021, Adhikary-2020}. What has not been sufficiently
explored is whether this multi-band feature combined with strong
correlation on Ni-$d$ orbitals may lead to some unique
experimental consequences in infinite-layer nickelates.

Recently a charge order (CO) with broken translational symmetry has
been reported independently by three experimental groups in the parent
compounds of nickelate superconductors
~\cite{Rossi-Lee-2112.02484,Kriger-Preziosi-2112.03341,Tam-Zhou-2112.04440}. This
raises a question concerning the CO origin and its competition with
other phases. CO in cuprates is well-known and has been intensively
studied~\cite{Comin-ARCMP-2016,Tranquada2020}.  Despite the
seeming similarity, the CO in nickelates has a few important
differences~\cite{Rossi-Lee-2112.02484,Kriger-Preziosi-2112.03341,Tam-Zhou-2112.04440}.
First, it already exists and has the highest onset temperature in
undoped compounds. Second, unlike cuprates where doped holes reside on
oxygen atoms, holes are mainly introduced into Ni-$d$ orbitals, as
supported by the experiments~\cite{Goodge2021,Rossi-prb2021}. Finally
and most importantly, it has an ordering wavevector $\mathbf{q}\simeq
\left(\frac{1}{3}, 0, 0\right)$ , compatible with a commersurate
modulation with period $3a_0$ ($a_0$ is the lattice constant). This
stripe pattern has not been observed in
cuprates~\cite{Tranquada-prb2011}. Hence, these differences may imply
a new mechanism for the CO in the nickelates, which is most probably
associated with the multi-band nature of the parent
compounds~\cite{Hepting-NatureMater2020,ZhangYangZhang-prb2020,Yang-Zhang2022,Wang-Zhang-Yang-Zhang,Hanghui-2020,Lechermann2020, Kang-2021, Hu-2019, Werner-2020,Zhang-2020,Wan-2021, Adhikary-2020}.



In this work, we first use density-functional-theory plus
dynamical-mean-field-theory (DFT+DMFT)
calculations~\cite{Hohenberg-PR-1964, Kohn-PR-1965, Georges-RMP-1996,
  Kotliar-RMP-2006} to reveal a special charge-transfer-driven CO mechanism in
a prototypical infinite-layer nickelate NdNiO$_2$. We find that for
every three Ni atoms, due to the presence of Nd-$d$/interstitial-$s$ derived
conduction bands near the Fermi level, electrons on one Ni atom are transferred
to the conduction bands under a reasonably large Hubbard interaction
$U_{\rm{Ni}}$ on Ni-$d$ orbitals, while electrons on the other two Ni
atoms become more localized. This leads to a characteristic
Ni$^{1+}$-Ni$^{2+}$-Ni$^{1+}$ CO stripe state.  Then we derive a
low-energy effective model based on the realistic band structure of
NdNiO$_2$. Using the effective model, we further elucidate that the
$\textbf{q}=(\frac{1}{3},0,0)$ CO stripe state arises from a
delicate competition between Hubbard interaction on Ni-$d$ orbitals
and charge transfer energy between Ni-$d$ orbitals and conduction
bands. Tuning the charge transfer energy controls not only stability of the CO
state but also its stripe pattern.
With Hubbard interaction strength and charge transfer energy in a
physically reasonable range, we find that the experimentally observed
$\mathbf{q}=\left(\frac{1}{3},0,0\right)$ stripe CO state is more stable than
both uniform paramagnetic state and usual checkerboard antiferromagnetic state.



\section*{Results}

\subsection*{DFT+DMFT calculations}

Unlike cuprate superconductors, concensus has not been
  reached regarding what is the minimal model to study infinite-layer
  nickelates. Therefore we first downfold the DFT-calculated band
  structure of NdNiO$_2$ to a large 17-orbital tight-binding model by using
  the maximally localized Wannier functions
  (MLWF)~\cite{Marzari2012}. The model includes five Ni-$d$ orbitals,
five Nd-$d$ orbitals, six O-$p$ orbitals and one interstitial-$s$
orbital per primitive cell. Our previous studies~\cite{Hanghui-2020,
  HanghuiChen-2022-Frontier} show that this 17-orbital model is
  sufficiently large that it can almost exactly reproduce the
  non-interacting band structure of NdNiO$_2$ in an energy window
  of about 15 eV around the Fermi level (see Supplementary Figure 1). A Slater-Kanamori interaction is added
on Ni-$d$ orbitals. We use DMFT to solve the interacting model. To
test a possible $\textbf{q}=\left(\frac{1}{3},0,0\right)$ CO state, we
use a $3\times1 \times 1$ supercell. Since in the 17-orbital model,
the MLWF for the Ni-$d$ orbitals are very localized and atomic-like
(see Supplementary Table 1), we use a Hubbard
$U_{\rm{Ni}}$ = 10 eV and a Hund's exchange $J_{\rm{Ni}}$ = 1 eV,
which has proved to be reasonable for this large energy window
treatment of transition metal oxides
\cite{Lechermann2020,L2020,Haule-PRB-2014,Haule-PRL-215}.  More
computational details are found in the Methods.

\begin{figure}[t!]
\centering
\includegraphics[width=0.95\linewidth]{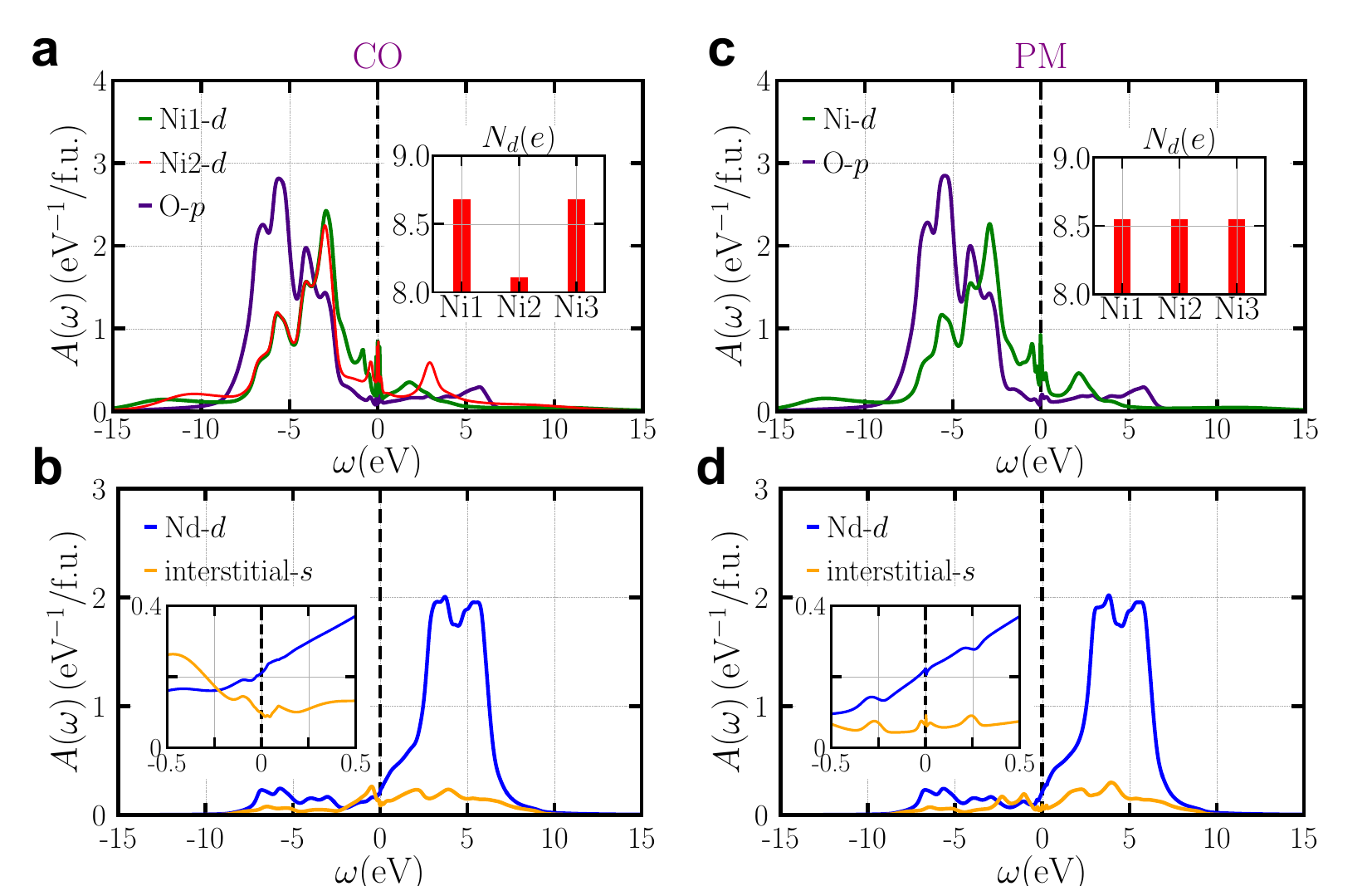}
\caption{\textbf{Spectral functions of NdNiO$_2$}. The spectral functions are
  calculated by DFT+DMFT method using the 17-orbital model with
  $U_{\rm{Ni}} = 10$ eV and $J_{\rm{Ni}} = 1$ eV. (a) and (b): The
charge ordered state (CO). (a): The green, red and purple curves are
Ni1-$d$, Ni2-$d$ and O-$p$ projected spectral functions,
respectively. The inset shows the site-resolved Ni-$d$ occupancy. The
$d$ occupancy of Ni2 site is substantially smaller than that of Ni1
and Ni3 sites. (b): The blue and orange curves are Nd-$d$ and
interstitial-$s$ projected spectral functions, respectively. The inset
shows the energy range near the Fermi level. (c) and (d): The uniform
paramagnetic state (PM). (c): The green and purple curves are Ni-$d$
and O-$p$ projected spectral functions, respectively. The inset shows
the site-resolved Ni-$d$ occupancy. (d): identical to (b) but for the
PM state. In all panels, the dashed line is the Fermi level, which is
shifted to zero point. Source data are provided as a Source Data file.}
\label{fig:17orb}
\end{figure}

We first add perturbations on the three Ni atoms in the simulation
cell.  After the self-consistency loop is converged, we do find a
charge-ordered (CO) state whose spectral function is shown in the
panels (a) and (b) of Fig.~\ref{fig:17orb}. For comparison, we also
enforce symmetry on the three Ni atoms and thus obtain a uniform
paramagnetic (PM) state whose spectral function is shown in the panels
(c) and (d). Panels (a) and (c) show the Ni-$d$ and O-$p$ projected
spectral functions. In the CO state, we find two different types of Ni
atoms. The spectral functions for Ni1-$d$ and Ni3-$d$ orbitals are
identical (collectively referred to as Ni1-$d$). Ni2-$d$ orbital is of
the other type. In the PM state, all the three Ni atoms are equivalent
due to the imposed symmetry.  The inset shows the Ni-$d$ occupancy. We
find that compared to Ni-$d$ occupancy in the uniform PM state,
  in the CO state the Ni1-$d$ occupancy is larger and approaches 9$e$, but the
Ni2-$d$ occupancy is substantially smaller and is close to 8$e$. This
leads to a characteristic Ni$^{1+}$-Ni$^{2+}$-Ni$^{1+}$ stripe
pattern. More importantly, the average Ni-$d$ occupancy in the CO
state is smaller than that in the PM state, indicating that the CO is
not a simple charge modulation between the three Ni atoms, but rather
there is a net charge transfer from the Ni2 atom to other atoms.
Panels (b) and (d) show the Nd-$d$ and interstitial-$s$ projected
spectral functions, which are almost identical between the CO and the PM
states. However, close to the Fermi level, we find appreciable
differences that Nd-$d$ and interstitial-$s$ states are more occupied
in the CO state than in the PM state. Therefore the emergence of CO state in NdNiO$_2$ is accompanied by a charge transfer from Ni2 atom
to Nd-$d$/interstitial-$s$ orbitals that form conduction bands
close to the Fermi level. We will show below that it is precisely
this charge transfer that can yield an energy gain and stabilize the
CO state.

\begin{figure}[t!]
\centering
\includegraphics[width=0.95\linewidth]{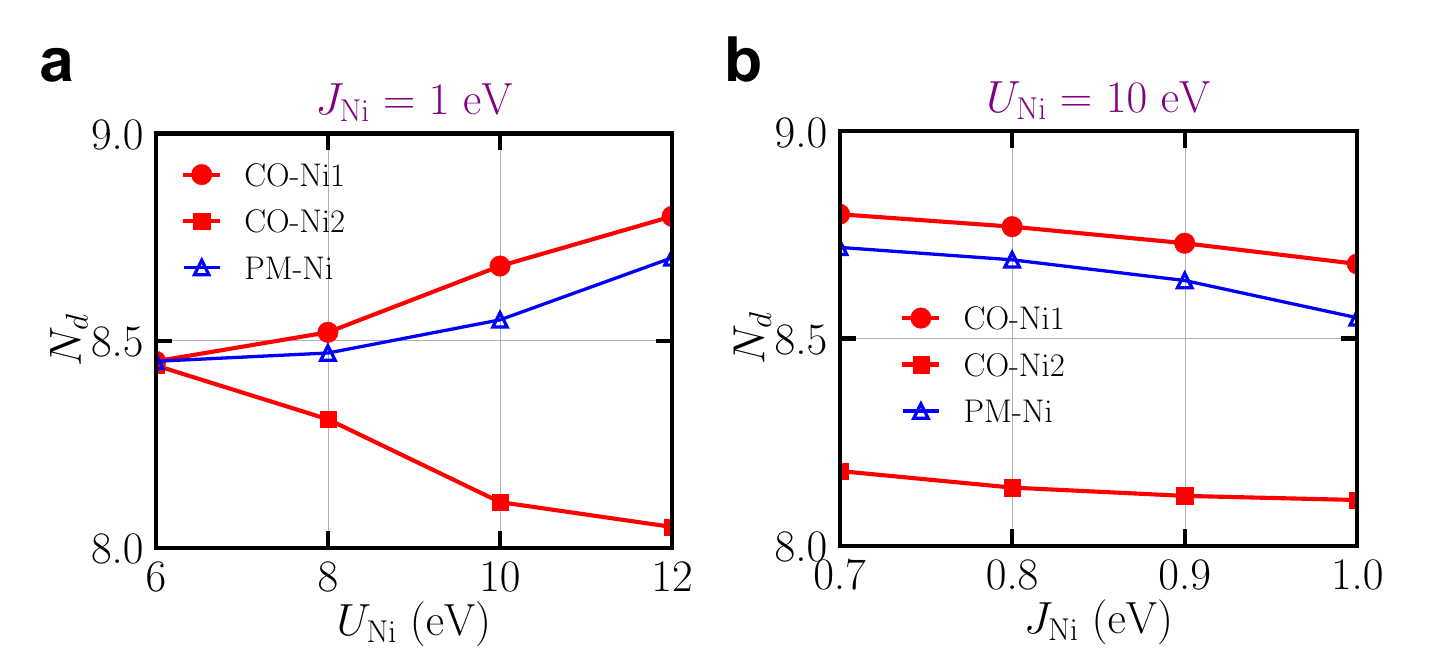}
\caption{\textbf{Dependence of Ni-$d$ occupancy $N_d$ on $U_{\rm{Ni}}$
    and $J_{\rm{Ni}}$}. The dependence of $N_d$ on $U_{\rm{Ni}}$ and
  $J_{\rm{Ni}}$ is calculated by using the 17-orbital model. (a):
  Ni-$d$ occupancy $N_d$ as a function of $U_{\rm{Ni}}$ with
  $J_{\rm{Ni}}$ = 1 eV. (b): Ni-$d$ occupancy $N_d$ as a function of
  $J_{\rm{Ni}}$ with $U_{\rm{Ni}}$ = 10 eV. In both panels, the red
  solid symbols and blue open symbols correspond to the Ni-$d$
  occupancy in the charge ordered (CO) and uniform paramagnetic (PM)
  states, respectively. Source data are provided as a
    Source Data file.}
\label{fig:UJdependence}
\end{figure}

Next we study how the charge disproportionation of the CO state
depends on $U_{\rm{Ni}}$ and $J_{\rm{Ni}}$. Panel (a) of
Fig.~\ref{fig:UJdependence} shows the Ni-$d$ occupancy of both CO and
PM states as a function of $U_{\rm{Ni}}$ with $J_{\rm{Ni}}$ being
fixed at 1 eV. As $U_{\rm{Ni}}$ exceeds 6 eV, the charge
disproportionation of Ni-$d$ occupancy develops in the CO state and
becomes more pronounced with $U_{\rm{Ni}}$. On the other hand,
changing $J_{\rm{Ni}}$ in a reasonable range (from 0.7 to 1.0 eV) with
$U_{\rm{Ni}}$ being fixed at 10 eV does not strongly affect the charge
disproportionation in the CO state. Furthermore we also test the
correlation effects from Nd-$d$ orbitals. By considering Hubbard
interaction on Nd-$d$ orbitals, we do not find qualitative changes in
the main results (see Supplementary Note 3).

\subsection*{Low-energy effective model}

\begin{figure}[t!]
\centering
\includegraphics[width=0.95\linewidth]{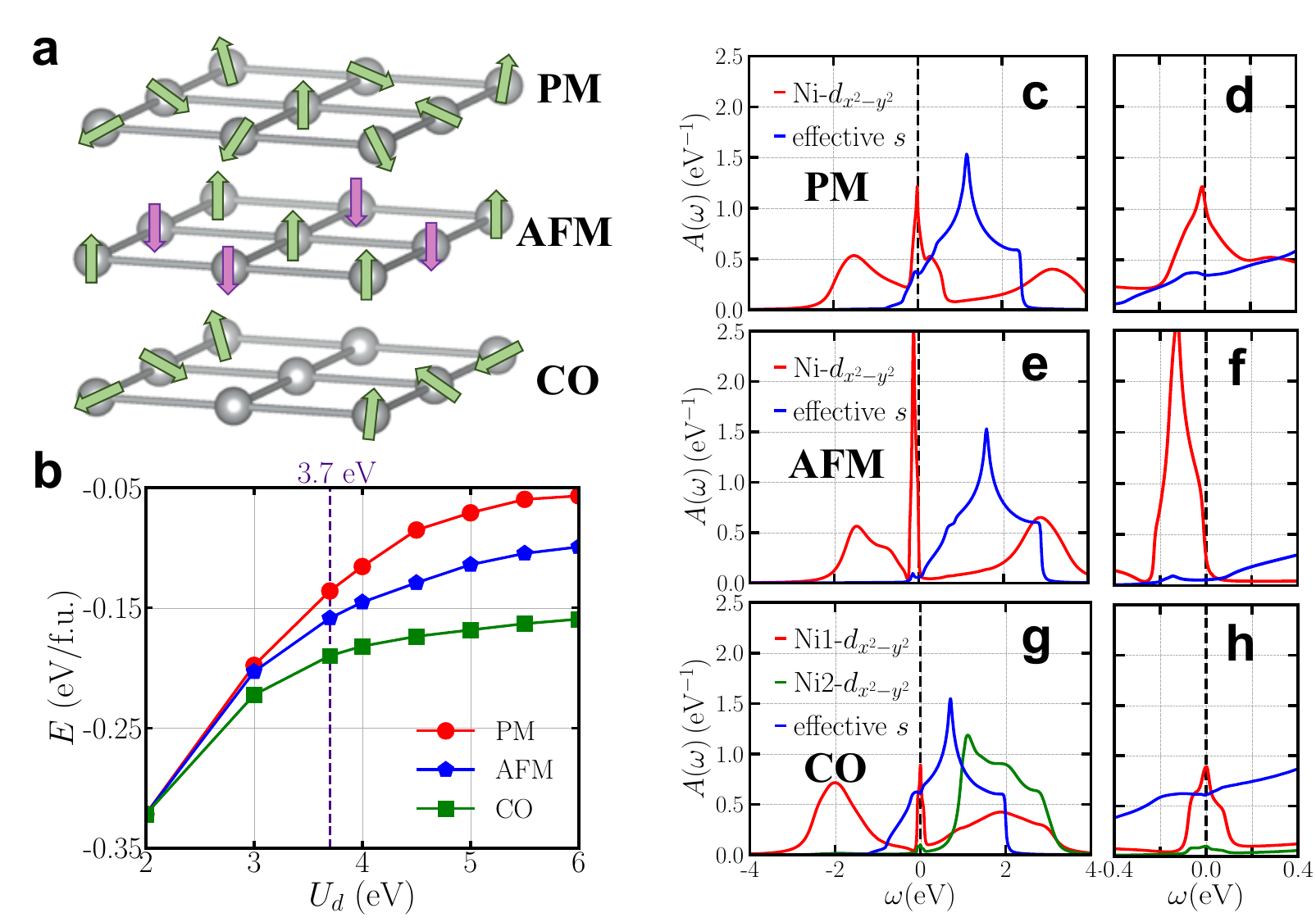}
\caption{\textbf{Low-energy effective model}.  (a): Schematics of the
  three competing states studied in the effective model: the uniform
  PM state (top), the checkerboard AFM state (middle) and the
  $\mathbf{q}=\left(\frac{1}{3},0,0\right)$ CO state (bottom). (b):
  The total energy $E$ calculated using the low-energy effective model
  Eq.~(\ref{eqn:model}) (up to a constant) in the PM (red), the AFM
  (blue) and the CO (green) states as a function of $U_d$. The purple
  dashed line highlights $U_d = 3.7$ eV, which reproduces the
  Ni-$d_{x^2-y^2}$ orbital effective mass from the \textit{ab initio}
  GW+EDMFT calculation~\cite{Petocchi-2020}. The error bar is smaller
  than the symbol size (see Methods).  (c)-(h): Orbital-resolved
  spectral functions calculated by using the low-energy effective
  model with $U_d = 3.7$ eV. (c): In the PM state. (e): In the AFM
  state. (g): In the CO state.  (d), (f) and (h) are the enlarged
  plots showing the corresponding near-Fermi-level features of the
  spectral function. Source data are provided as a Source
    Data file.}
\label{fig:spectral}
\end{figure}

To gain a deeper understanding of origin and generality of the CO
state that is found in the DFT+DMFT calculations of NdNiO$_2$, we
build a low-energy effective model. We want to make the model simple,
which aims to qualitatively reproduce the key features of the CO
state. The advantage of such a simple effective model is
  that we can obtain an accurate self-energy so as to resolve the
  total energy difference between different competing phases.  In
  addition, it also helps us identify the control parameters for the
  CO state. The effective model, equipped with proper energy
  dispersions, may also be applid to the study of charge ordering
  phenomenon in other strongly correlated materials. From the
preceding DFT+DMFT calculations, we find that there are two key
ingredients in the formation of the CO state in NdNiO$_2$. One is the
Hubbard interaction on Ni-$d$ orbitals and the other is the charge
transfer from Ni-$d$ orbitals to the conduction bands that consist of
Nd-$d$/interstitial-$s$ orbitals.  In order to describe the interplay
between these two factors, we include a correlated orbital and an
auxiliary orbital in the effective model. To make connection to
infinite-layer nickelates, we choose Ni-$d_{x^2-y^2}$ orbital as the
correlated orbital and add local Hubbard interaction on it. We also
choose an effective-$s$ orbital as the auxiliary orbital that provides
a free conduction band. We note that in real NdNiO$_2$, there are
multiple conduction bands that are composed of Nd-$d$ and
interstitial-$s$ orbitals. Here the effective-$s$ orbital (not to be
confused with the interstitial-$s$ orbital) is employed to simplify
the description of conduction bands. Similar models of a correlated
Ni-$d_{x^2-y^2}$ orbital plus free conduction bands have been used to
study the low-energy physics of infinite-layer
nickelates~\cite{Hanghui-2020, wu2019robust, Adhikary-2020,
  Nomura-prb2019, Plienbumrung-PRB-2022,Jiang-2022,Peng-2021}. The
full Hamiltonian of our effective model takes the form:
\begin{equation}
\label{eqn:model} \hat{H}  = \sum_{\textbf{k} \sigma} \left[\epsilon_d(\textbf{k})\hat{d}^{\dagger}_{\textbf{k}\sigma}\hat{d}_{\textbf{k}\sigma} + \epsilon_s(\textbf{k})\hat{s}^{\dagger}_{\textbf{k}\sigma}\hat{s}_{\textbf{k}\sigma} + \big(V_{ds}(\textbf{k})\hat{d}^{\dagger}_{\textbf{k}\sigma}s_{\textbf{k}\sigma} + \rm{h.c.}\big)\right] + U_d\sum_i \hat{n}^d_{i\uparrow}\hat{n}^d_{i\downarrow} - \hat{V}^{\rm{dc}}
\end{equation} 
where $\hat{d}^{\dagger}_{\textbf{k}\sigma}$
($\hat{s}^{\dagger}_{\textbf{k}\sigma}$) is the creation operator on
Ni-$d_{x^2-y^2}$ (effective-$s$ orbital) with momentum $\textbf{k}$
and spin $\sigma$.  $\hat{n}^d_{i\sigma} =
\hat{d}^{\dagger}_{i\sigma}\hat{d}_{i\sigma}$ is the occupancy
operator of Ni-$d_{x^2-y^2}$ orbital at site $i$ with spin $\sigma$.
$\hat{V}^{\rm{dc}}$ is the double counting
potential~\cite{Ylvisaker-2009}.  The details of
Eq.~(\ref{eqn:model}) are found in the Methods. In order to study the
CO found in NdNiO$_2$, we fit all the hopping parameters in the energy
dispersion and hybridization to the DFT band structure. The
non-interacting part of Eq.~(\ref{eqn:model}) reproduces the
near-Fermi-level DFT band structure of NdNiO$_2$, in which there are
two bands crossing the Fermi level (see Supplementary Figure 1). We mention one important parameter in the
energy dispersion: the bare onsite energy difference between
Ni-$d_{x^2-y^2}$ and effective-$s$ orbitals $E_{ds} = 0.70$ eV (see
the Methods), which we will show is another key control parameter for
the CO state besides the Hubbard interaction on Ni-$d$ orbitals.

Next we consider three competing states in the low-energy effective
model (see Figure~\ref{fig:spectral}(a) for a schematic): the uniform
PM state with dynamically fluctuating spins (top), the checkerboard
AFM (middle) and the CO state with an ordering wavevector
$\mathbf{q}=\left(\frac{1}{3},0,0\right)$ (bottom).
Due to simplicity of the model
  Eq.~(\ref{eqn:model}), we can obtain a highly accurate self-energy
  of Ni $d_{x^2-y^2}$ orbital, which enables us to calculate the total
  energy of each state as a function of $U_d$ and resolve the total
  energy difference between various competing phases (the details of
  the total energy calculations are found in the Methods).  The
results are shown in Fig.~\ref{fig:spectral}(b). We find that when
$U_d$ exceeds a critical value, the AFM and CO states can both be
stabilized. More importantly, the CO state has lower energy than the
AFM state and the PM state.  To find a proper $U_d$ for our model, we
calculate the effective mass of the Ni-$d_{x^2-y^2}$ orbital (see Supplementary
Note 4 for details). We find that $U_d$ = 3.7 eV
leads to an effective mass of 5.5, which exactly reproduces the result
from the full \textit{ab initio} GW+EDMFT
calculations~\cite{Petocchi-2020}. At this $U_d$ value, the CO state
has a lower energy than the PM state by about 50 meV/f.u., a magnitude
reasonably consistent with the CO onset of infinite-layer nickelates
in
experiment~\cite{Rossi-Lee-2112.02484,Kriger-Preziosi-2112.03341,Tam-Zhou-2112.04440}. Furthermore,
the CO state is also more stable than the AFM state by about 30
meV/f.u. which may provide an explanation for the lacking of
long-range AFM ordering in infinite-layer nickelates despite the
substantial AFM superexchange found in experiment~\cite{Lu-Science2021,
  Zhao-2021}.

Figure~\ref{fig:spectral}(c)-(h) compares the spectral functions of
the low-energy effective model Eq.~(\ref{eqn:model}) for the PM, AFM
and CO states (calculated with $U_d = 3.7$ eV).
Fig.~\ref{fig:spectral}(c) displays the orbital-resolved spectral
functions of the PM state. The red (blue) curve is the
Ni-$d_{x^2-y^2}$ (effective-$s$) projected spectral
function. Ni-$d_{x^2-y^2}$ orbital exhibits a characteristic
three-peak feature: lower Hubbard band (LHB), upper Hubbard band (UHB)
and a quasi-particle peak around the Fermi level. The effective-$s$
orbital has an occupancy of about 0.1$e$/f.u., due to the self-doping
effect. This result is consistent with the previous studies using
similar effective models~\cite{Hanghui-2020, Karp-2020-PRX,
  Botana-Norman-prx2020}. Fig.~\ref{fig:spectral}(d) zooms-in the
Ni-$d_{x^2-y^2}$ projected spectral function close to the Fermi level,
which highlights the quasi-particle peak and the partially occupied
conduction band. Fig.~\ref{fig:spectral}(e) and \ref{fig:spectral}(f)
show the spectral function of the AFM state. Compared to the PM state,
the quasi-particle peak is more pronounced but shifts away from the
Fermi level. The self-doping effect is weaker in the AFM state.
Fig.~\ref{fig:spectral}(g) shows the orbital-resolved spectral
functions of the CO state, where the red and green curves are for the two
different types of Ni-$d_{x^2-y^2}$ orbital, and the blue curve
is for the effective-$s$ orbital. Similar to the PM state, the
Ni1-$d_{x^2-y^2}$ orbital in the CO state (red curve) exhibits a
three-peak feature, but the Ni LHB moves towards lower energy.
This indicates that the Ni1-$d_{x^2-y^2}$ orbital becomes more
localized (i.e. closer to half-filling) in the CO state. In contrast,
the Ni2-$d_{x^2-y^2}$ orbital in the CO state is almost
empty. Compared to the PM state, the effective-$s$ orbital is more
occupied. The spectral functions in panel (g) qualitatively reproduces the
Ni$^{1+}$-Ni$^{2+}$-Ni$^{1+}$ stripe pattern and the charge
transfer from Ni2 site to the conduction bands, which are found from the
preceding DFT+DMFT calculations. Panel (h) shows an enlarged plot of the
Ni-$d_{x^2-y^2}$ projected spectral functions near the Fermi level in
the CO state. The Ni-$d_{x^2-y^2}$ quasi-particle peak in the CO state
has a smaller peak value than in the PM state (calculated at the same
$U_d$), indicating that in addition to local Hubbard interaction, the
formation of long-range CO further suppresses the quasi-particle peak
(the effect is more pronounced at larger $U_d$, see Supplementary Note 6).

\subsection*{Energetics}

\begin{figure}[t!]
\centering
\includegraphics[width=0.9\linewidth]{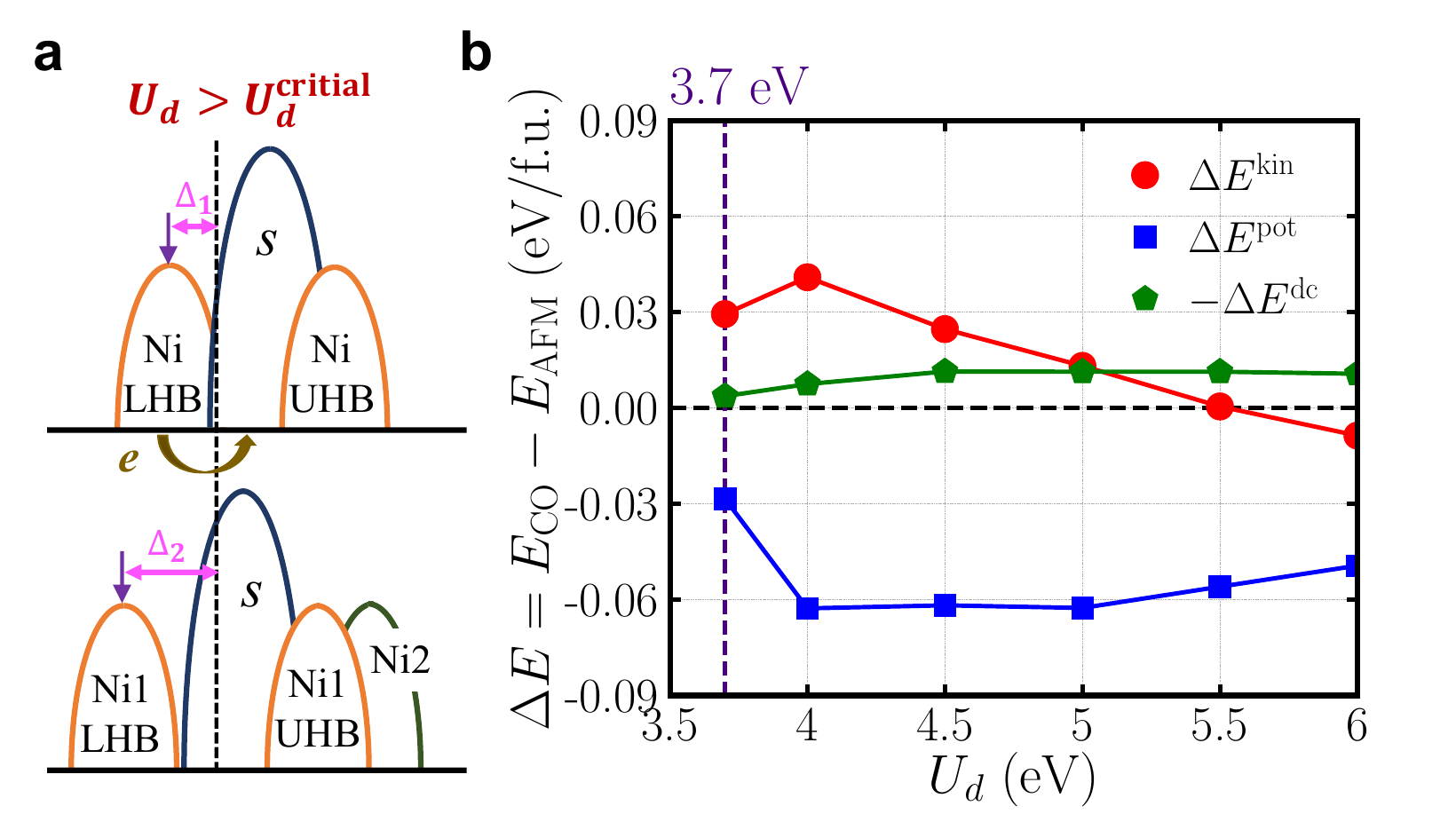}
\caption{\textbf{Charge transfer and energetics}. (a): Schematics of
  the interacting electronic structure of the low-energy effective
  model Eq.~(\ref{eqn:model}). The brown arrow indicates the charge
  transfer from a Ni site to neighboring effective-$s$ orbitals. The
  upper panel is the one before the charge transfer and the lower
  panel is after the charge transfer (when
  $U_d>U^{\rm{critical}}_d$). (b): The total energy difference between
  the CO and the AFM states decomposed into kinetic energy (kin,
  red circles), potential energy (pot, blue
    squares) and double-counting energy (dc, green
    pentagons) as a function of $U_d$, calculated by using the
  low-energy effective model Eq.~(\ref{eqn:model}). The purple dashed
  line highlights $U_d = 3.7$ eV.  The error bar is smaller than the
  symbol size (see Methods). Source data are provided as
    a Source Data file.}
\label{fig:gain}
\end{figure}

Next we analyze the energetics of the low-energy
effective model in more details. In particular, we study why the CO
state may have a lower energy than the AFM
state. Figure~\ref{fig:gain}(a) shows a schematic electronic structure
of the effective model Eq.~(\ref{eqn:model}). The brown arrow
indicates the charge transfer from a Ni site to neighboring
effective-$s$ orbitals. The upper panel is the one before the charge
transfer, while the lower panel is after the charge transfer (when
$U_d$ exceeds the critical value). The CO can emerge if there is an
overall energy gain from the charge transfer. However, the energy of
Ni LHB is lower than the effective-$s$ orbitals. Thus the electron
transfer from the Ni2 site to the effective-$s$ orbitals seems to
result in an energy cost. To solve this paradox, we compare the total
energy of the CO and AFM states and decompose the energy difference
into three contributions: $E = E^{\mathrm{kin}} +
E^{\mathrm{pot}}-E^{\mathrm{dc}}$, where $E^{\mathrm{kin}}$ is the
kinetic energy, $E^{\mathrm{pot}}$ is the potential energy and
$E^{\mathrm{dc}}$ is the double-counting energy (see the Methods).

We first analyze the case of $U_d = 3.7$ eV, as highlighted by the
purple dashed line in Figure~\ref{fig:gain}(b). We find that compared
to the AFM state which does not have charge transfer, the charge
transfer indeed results in an energy cost in $E^{\rm{kin}}$ for the CO
state. This is consistent with the schematics shown in
Fig.~\ref{fig:gain}(a).  However, the charge transfer also leads to an
energy gain in $E^{\rm{pot}}$. That is because in the AFM state, every
Ni site has almost one electron on the $d_{x^2-y^2}$ orbital.  Due to the
repulsive Hubbard interaction,
$E^{\mathrm{pot}}=\frac{1}{2}\text{Tr}(\hat{\Sigma}_{\text{cor}}\hat{G}_{\text{cor}})>
0$ for each Ni. In the CO state, the Ni1 and Ni3 sites have one
electron on the $d_{x^2-y^2}$ orbital. But the charge transfer
depletes the electron on the Ni2-$d_{x^2-y^2}$ orbital, so that its
self-energy $\Sigma_{\mathrm{cor}}$ essentially becomes zero. This
decreases the potential energy and always leads to an energy gain
$\Delta E^{\mathrm{pot}} < 0$ when $U_d$ exceeds the critical value.
As for the double-counting energy,
$E^{\text{dc}}=\frac{1}{2}U_dN_d(N_d-1)$~\cite{Czyzyk1994}. The AFM
state is close to a half-filled configuration for every Ni site, while the
CO state has a configuration in which Ni2 site has almost zero occupancy
and Ni1/Ni3 site is half-filled. Therefore, in both states, the double
counting energy $E^{\rm{dc}}$ is small and so is their difference
(compared to $\Delta E^{\rm{pot}}$). In short, there is a delicate competition
between charge transfer energy and local Hubbard interaction. When the
potential energy gain outweighs the kinetic energy cost during the
charge transfer, the CO state becomes more stable than the AFM
state.

Then we analyze the case of a general $U_d$ (see
Figure~\ref{fig:gain}(b)).  We find that the trend does not change in
$\Delta E^{\rm{pot}}$ and $\Delta E^{\rm{dc}}$. However, when $U_d$ is
sufficiently large, $\Delta E^{\rm{kin}}$ itself becomes negative
(i.e.  an net energy gain after the charge transfer). That is because as
$U_d$ increases, the Ni1/Ni3 LHB moves towards the lower energy after
the charge transfer (i.e. $|\Delta_2| > |\Delta_1|$). This effect
decreases the kinetic energy and eventually counteracts the kinetic
energy cost from the charge transfer (brown arrow in
Fig.~\ref{fig:gain}(a)).

\subsection*{CO stability and stripe pattern}

\begin{figure}[t!]
\centering
\includegraphics[width=0.95\linewidth]{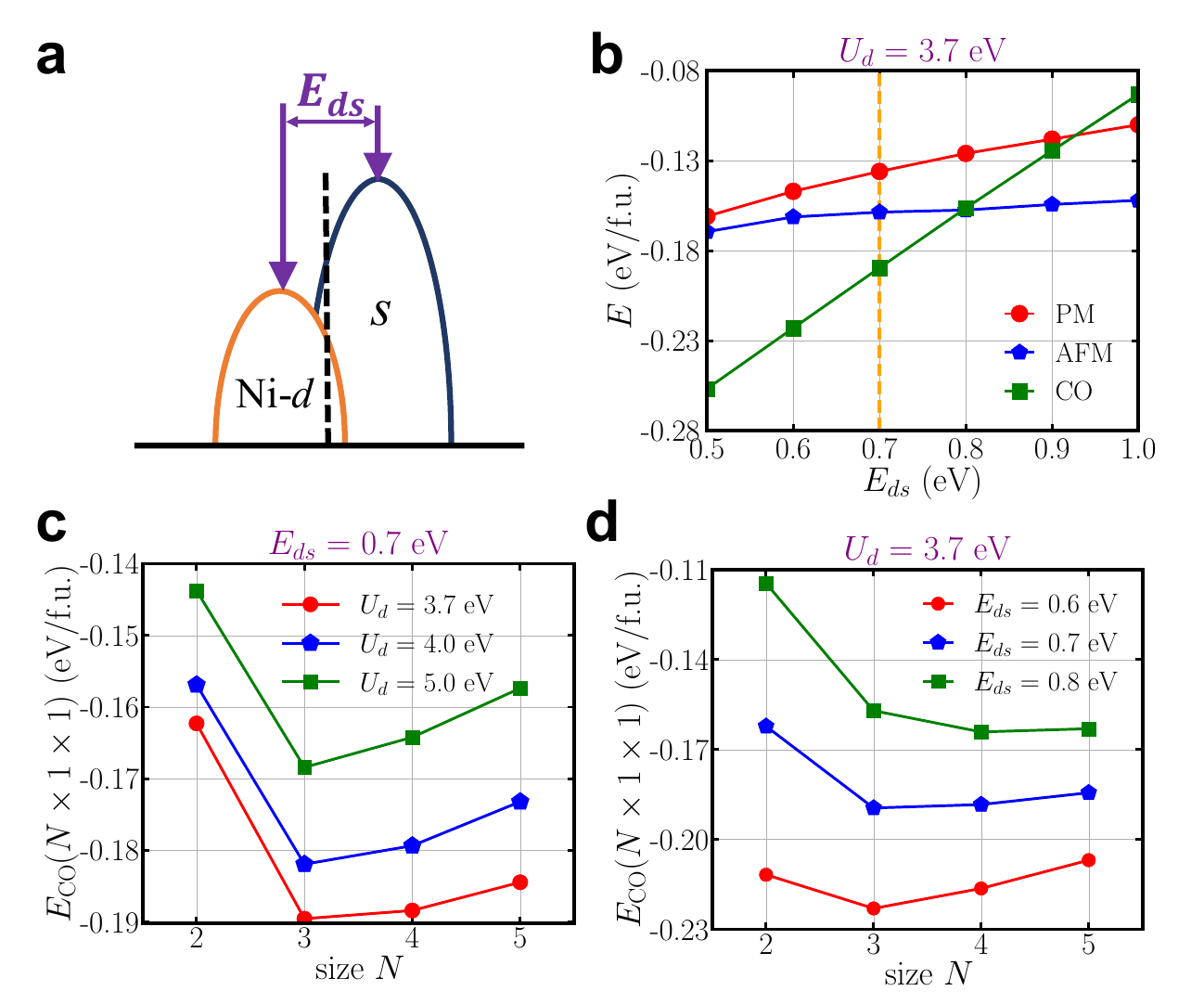}
\caption{\textbf{CO stability and stripe pattern}. (a): Schematics of
  the non-interacting electronic structure of the low-energy effective
  model Eq.~(\ref{eqn:model}) where $E_{ds}$ is the energy separation
  between the bare energy of Ni-$d_{x^2-y^2}$ and effective-$s$
  orbitals. (b): The total energy (up to a constant) of the PM (red),
  the AFM (blue) and the CO (green) states as a function of $E_{ds}$,
  calculated by using the low-energy effective model
  Eq.~(\ref{eqn:model}) with $U_d = 3.7$ eV.  The orange dashed line
  highlights $E_{ds} = 0.7$ eV, which is obtained from fitting to the
  DFT band structure of NdNiO$_2$. (c): The total energy of different
  stripe CO states with the wavevector
  $\textbf{q}=\left(\frac{1}{N},0,0\right)$ as a function of $U_d$,
  calculated by using the low-energy effective model
  Eq.~(\ref{eqn:model}) with $E_{ds} = 0.7$ eV. (d): The total energy
  of different stripe CO with the wavevector
  $\textbf{q}=\left(\frac{1}{N},0,0\right)$ as a function of $E_{ds}$,
  calculated by using the low-energy effective model with $U_d = 3.7$
  eV.  In panels (b), (c) and (d), the error bar is smaller than the
  symbol size (see Methods). Source data are provided as
    a Source Data file.}
\label{fig:CT}
\end{figure}

The preceding analysis shows that the charge transfer from
Ni-$d_{x^2-y^2}$ orbital to the near-Fermi-level conduction bands
plays a crucial role in the formation of the CO state. In our
effective model, the conduction bands are derived from the
effective-$s$ orbitals. Figure~\ref{fig:CT}(a) is a schematic
non-interacting electronic structure of the effective model
Eq.~(\ref{eqn:model}). We show below that the charge transfer energy
$E_{ds}$, which is defined as the energy separation between
the bare onsite energy of Ni-$d_{x^2-y^2}$ and effective-$s$ orbitals,
controls not only stability but also stripe pattern of the CO
state in the effective model.

We first examine the total energy of the PM, AFM and CO states as a
function of $E_{ds}$ using the effective model
Eq.~(\ref{eqn:model}). For each $E_{ds}$, we recalculate the total
energy of each state and show the results in
Figure~\ref{fig:CT}(b). We use $U_d = 3.7$ eV but find a similar trend
using other values of $U_d$ (see Supplementary Note 7).  Fitting to
the DFT band structure, we obtain $E_{ds} = 0.7$ eV (the orange dashed
line in Fig.~\ref{fig:CT}(b)), at which the CO state is more stable
than the AFM and PM states.  As $E_{ds}$ increases, the total energy
of the CO state first becomes higher than the AFM state ($E_{ds} >
0.8$ eV) and then even higher than the PM state ($E_{ds} > 0.9$
eV). This can be understood because increasing $E_{ds}$ leads to a
larger kinetic energy cost in the charge transfer (as we demonstrated
in Fig.~\ref{fig:gain}(a)).



Next we study different stripe CO patterns and compare their total
energies using the effective model Eq.~(\ref{eqn:model}). We use a
$N\times 1 \times 1$ supercell to study the stripe CO with the
wavevector $\textbf{q}=\left(\frac{1}{N},0,0\right)$. Before
  discussing the results, we first present an intuitive picture of how
  the CO total energy may depend on $N$. As we have shown, the CO
  state emerges from a charge transfer of almost one electron from one
  Ni site to the conduction bands for every $N$ Ni atoms. During the
  charge transfer process, the electron needs to overcome the charge
  transfer energy $E_{ds}$. This results in a kinetic energy cost
  $\propto E_{ds}$. On the other hand, the charge transfer depletes
  the charges on that Ni site. Since the local Hubbard interaction is
  repulsive in our model, this leads to a potential energy gain
  $\propto U_d$. Thus the total energy of the entire system
  ($N$-Ni-atom cell) is:
\begin{equation}
  \label{eq1XX} E_{N\textrm{-Ni-atom}} = \alpha E_{ds} - \beta U_d + N E_0
\end{equation} 
where $\alpha$ and $\beta$ are both positive and $E_0$ is the energy contribution that does not change during the charge transfer. The tricky point is that both $\alpha$ and $\beta$ also depend on the wavevector or the wavelength (i.e. $N$). Therefore the total energy per-Ni-atom cell is:
\begin{equation}
\label{eq2XX} E_{\textrm{per-Ni-atom}} = \frac{E_{N\textrm{-Ni-atom}}}{N}=\alpha(N) \frac{E_{ds}}{N} - \beta(N) \frac{U_d}{N} + E_0
\end{equation} For given $E_{ds}$ and $U_d$, minimizing $E_{\textrm{per-Ni-atom}}$ with respect to $N$ yields the optimal wavelength $N$ or equivalently optimal wavevector $\textbf{q}$. However, due to interaction, the analytical expressions of $\alpha(N)$ and $\beta(N)$ are not known. Therefore we numerically calculate $E_{N\textrm{-Ni-atom}}$ as a function of $N$ for different $E_{ds}$ and $U_d$.

Figure~\ref{fig:CT}(c) compares the total energy of different stripe
CO states for a few representative $U_d$ at $E_{ds}$ = 0.7 eV (DFT-fitted
value). We find that at $E_{ds} = 0.7$ eV, the
$\textbf{q}=\left(\frac{1}{3},0,0\right)$ CO has the lowest total
energy in a range of $U_d$ values. The reason that the total energy of
the stripe CO state does not monotonically change with $N$ is because
with $N$ increasing, on one hand, the kinetic energy cost decreases;
on the other hand, the potential energy gain also decreases. These two
factors are competing and the calculations find that the total energy
minimum appears at an optimal CO wavevector
$\textbf{q}=(\frac{1}{3},0,0)$. While our model is simplified, this
result is consistent with the experimentally observed CO stripe
pattern of infinite-layer
nickelates~\cite{Rossi-Lee-2112.02484,Kriger-Preziosi-2112.03341,Tam-Zhou-2112.04440}. Fig.~\ref{fig:CT}(d)
compares the total energy of different stripe CO states for a few
representative $E_{ds}$ at $U_{d}$ = 3.7 eV. We find that as $E_{ds}$
increases from 0.6 eV to 0.8 eV, the wavevector of the lowest energy
CO state evolves from $\left(\frac{1}{3},0,0\right)$ to
$\left(\frac{1}{4},0,0\right)$. This is because as $E_{ds}$ changes,
the decreasing rate of the kinetic energy with $N$ also changes, which
leads to a different optimal wavevector for the CO
state. We mention that when $E_{ds} \geq 0.8$ eV, the AFM
  state has lower energy than the CO state (see
  Fig.~\ref{fig:CT}(b)). This means that based on our model
  calculations, the CO state with a wavevector of
  $\textbf{q}=\left(\frac{1}{4}, 0, 0\right)$ can not be
  observed. Instead, it gives way to the AFM state. To summarize, in
the phase space spanned by $(U_d, E_{ds})$, our effective model
Eq.~(\ref{eqn:model}) has a parameter range in which the stripe CO
state is more stable than the AFM and PM states with an optimal
wavevector $\textbf{q}=(\frac{1}{3},0,0)$, which reproduces the
experimental observations of infinite-layer
nickelates~\cite{Rossi-Lee-2112.02484,Kriger-Preziosi-2112.03341,Tam-Zhou-2112.04440}.

\section*{Discussion}
  
We make a few comments before conclusion.

First, we note that CO or charge density wave (CDW) may have different
origins. In addition to electron-electron interaction, electron-phonon
interaction may also play an important role in the formation of CDW,
which has been intensively studied in other material systems such as
transition metal dichalcogenides~\cite{Manzeli2017}. A CDW state that
arises from electron-phonon interaction is accompanied by a structural
distortion. Thus, in addition to using the ideal lattice structure of
infinite-layer nickelates, we also study possible structural
distortions in the CO state~\cite{Hanghui-2022,Huang-2022}. Since the
charge transfer breaks the original translational symmetry, oxygen
atoms may move away from their ideal positions (see Supplementary Note
5). The movement of oxygen atoms can be modelled by changing the
on-site energy of Ni-$d_{x^2-y^2}$ orbital in the effective model
Eq.~(\ref{eqn:model}). Our calculations find that the oxygen atom
movements are minute ($\sim$0.05~\AA), which is in agreement with the
other study ~\cite{ZhangRuiqi-2022}. The resulting energy gain in CO
is less than 5 meV/f.u., about one order of magnitude smaller than the
energy difference shown in Fig.~\ref{fig:spectral}(b). The energy gain
may be used as a rough estimation to quantify the relative importance
of different mechanisms. Our results imply that the CO state found in
infinite-layer nickelates is more likely of electronic origin, while
structural distortions driven by electron-phonon interaction may play
a secondary role.

Second, the CO state from our calculations has a substantial charge
disproportionation, driven by a charge transfer from a portion of Ni
atoms to the near-Fermi-level conduction bands. A direct
  consequence is that the average Ni occupancy in the CO state is
  substantially smaller than that of uniform paramagnetic state. Such
  a CO state is dissimilar to the charge density wave state found in
  the calculations of doped single-band Hubbard
  model~\cite{JiangYifan-PRResearch-2020,zheng2017stripe}, doped
  $t$-$J$ model~\cite{JiangHongchen-PRB-2018} and a 2-orbital model of
  Ref.~\cite{Peng-2021}, in which the charge profile shows a Friedel
  oscillation with the average occupancy per site being equal to that
  of uniform paramagnetic state. In those states, variation in site
  occupancy is due to charge transfer between neighboring correlated sites.

Third, as we mentioned previously, the analytical expression of CO
energy dependence on wavevector $E(\textbf{q})$ is not
known. Therefore we can not find the optimal wavevector \textbf{q} by
solving $\nabla_{\textbf{q}} E(\textbf{q})=0$. Instead we have to test
as many known competing states as possible (within computational
capability). In addition to the
$\textbf{q}=\left(\frac{1}{N},0,0\right)$ CO state, we also test a
different CO state which can be accommodated by the available
simulation cell (see Supplentary Note 8).  We use the low-energy
effective model to study all these competing states (PM state, AFM
state and five CO states with different wavevectors) and compare their
total energy. Our calculations find that using the DFT-fitted energy
dispersion, for a reasonable range of $U_d$, the
$\textbf{q}=\left(\frac{1}{3},0,0\right)$ CO state is the most
energetically favorable one among all the states considered.

Fourth, a number of
studies~\cite{Wan-2021,Wang-2020,Kang-2021,Kreisel-PRL-2022} have
shown that Ni $d_{3z^2-r^2}$ orbital plays an important role in
infinite-layer nickelates. Comparison of the low-energy effective
model calculations to the 17-orbital model calculations which include
all five Ni-$d$ orbitals reveals that the minimum physics needed to
account for the CO phenomenon is a charge transfer from a correlated
Ni-$d_{x^2-y^2}$ orbital to a conduction band that is close to the
Fermi level. Adding other orbitals into the effective model may
describe more physical phenomena of infinite-layer nickelates, but as
far as charge ordering is concerned, the 2-orbital effective model is
sufficient.

Finally, we briefly discuss how one may experimentally
  distinguish CO states of different origins in infinite-layer nickelates. If
  the CO in infinite-layer nickelates arises from the revealed
  charge-transfer mechanism, it will exhibit a mixed valence state of
  Ni, a characteristic Ni$^{1+}$-Ni$^{2+}$-Ni$^{1+}$ stripe pattern,
  as well as a unidirectional Ni$^{2+}$ chain in which holes are
  localized. In experiment, one may probe this mixed valence state of
  Ni either by x-ray absorption spectroscopy (XAS) or cross-sectional
  electron energy loss spectroscopy (EELS)~\cite{Comin-ARCMP-2016}.
  XAS can measure the average Ni valence, whose spectrum in the charge
  ordered state should be distinct from that of pure Ni$^{1+}$ or
  Ni$^{2+}$. A more informative measurement is the cross-sectional
  EELS, which may provide a spatially resolved spectrum of Ni$^{1+}$
  and Ni$^{2+}$ columns in the charge ordered state. In addition,
  scanning tunneling microscopy (STM) can image the localized holes in
  the unidirectional Ni$^{2+}$ chain~\cite{Comin-ARCMP-2016}.  On the
  other hand, if the CO or charge density wave (CDW) is driven by
  phonon and electron-phonon coupling, a characteristic phonon of the
  ideal crystal structure should become soft around the CDW transition
  temperature and the emergence of CDW accompanies a
  periodic-lattice-distortion, which lowers the total
  energy~\cite{Manzeli2017,Holt-PRL-2001,Weber-PRL-2011}. The
  vibrational mode of characteristic phonon should match the pattern
  of periodic-lattice-distortion observed at low
  temperatures~\cite{Holt-PRL-2001,Weber-PRL-2011}.  The magnitude of
  the periodic-lattice-distortion should be such that the resulting
  energy gain is comparable to the CDW transition
  temperature~\cite{Manzeli2017}. In experiment, inelastic x-ray
  scattering or neutron scattering can measure temperature dependence
  of phonon dispersion and identify soft phonons, if they exist around
  CDW temperature~\cite{Weber-PRL-2011}. The low-temperature
  perodic-lattice-distortion can be probed either by x-ray diffraction
  (XRD) or by transmission electron microscopy (TEM). Emergence of
  perodic-lattice-distortion lowers the crystal symmetry and thus
  changes the diffraction pattern in XRD
  measurements~\cite{Holder-ACSNano-2019}. TEM, in particular with the
  recent developments of electron
  ptychography~\cite{Chen-Science-2021}, can achieve atomic-resolution
  imaging, enabling visualization of perodic-lattice-distortion in
  real space and accurate measurements of structural
  distortions. Using the lattice distortions found from TEM
  measurements, one can estimate the energy gain from the
  periodic-lattice-distortion by performing first-principles calculations and
  then compare the energy gain to the experimental CDW transition
  temperature. In short, the experimental observation of ``a mixed
  valence state of Ni/localized holes on unidirectional Ni$^{2+}$
  chain'' versus ``soft phonon/periodic-lattice-distortion'' may
  distinguish different origins of charge order in infinite-layer
  nickelates. We hope that the two sets of experiments outlined above,
  as well as other experimental techniques, will identify the true
  underlying mechanism among various theoretical proposals.


To conclude, our first-principles calculations show that a
$\mathbf{q}=\left(\frac{1}{3},0,0\right)$ CO state, which has a
characteristic Ni$^{1+}$-Ni$^{2+}$-Ni$^{1+}$ stripe pattern, is
stabilized in infinite-layer nickelates when Hubbard interaction
on Ni-$d$ orbitals is reasonably large. For every three Ni atoms, the
CO state is driven by a charge transfer from one Ni atom to
near-Fermi-level conduction bands, leaving electrons on the other two
Ni atoms to become more localized. We further derive a simple
low-energy effective model and elucidate that stabilization of the
CO state over PM and AFM states is the consequence of a delicate
competition between Hubbard interaction $U_d$ on a correlated
orbital and charge transfer energy $E_{ds}$ between the correlated
orbital and the conduction band. Such a competition may induce CO
states in other correlated materials besides infinite-layer
nickelates. Manipulating the CO state in infinite-layer nickelates by
tuning the charge transfer energy may also affect their
superconductivity~\cite{Alvarez-arXiv-2022},
which deserves further experimental study.

\section*{Methods}

\subsection*{DFT+DMFT calculations}

We perform density functional theory (DFT) plus dynamical mean field theory
(DMFT) calculations~\cite{Hohenberg-PR-1964, Kohn-PR-1965,
  Georges-RMP-1996, Kotliar-RMP-2006} with the aid of maximally
localized Wannier functions (MLWF)~\cite{Marzari2012,mostofi2008wannier90}.

The DFT method is implemented in the Vienna ab initio simulation
package (VASP) code~\cite{Kresse1996} with the projector augmented
wave (PAW) method~\cite{kresse1999}. The Perdew-Burke-Ernzerhof
(PBE)~\cite{perdew1996} functional is used as the exchange-correlation
functional in DFT calculations. The Nd-$4f$ orbitals are treated as
core states in the pseudopotential because the $4f$ shell
  is hidden in the core of the Nd ion, and Hubbard interaction on $4f$
  orbitals is very large ($> 10$ eV). As a result, the Nd-$4f$
  electrons are far from the Fermi level and turn into strongly
  localized $f$ spins. The direct coupling between Nd-$4f$ electrons
  and itinerant electrons is minute~\cite{Been-2021}.
  We use an energy cutoff of 600
eV and sample the Brillouin zone by using $\Gamma$-centered
\textbf{k}-mesh of $12\times12\times12$ per primitive cell. The
crystal structure is fully relaxed with an energy convergence
criterion of $10^{-6}$ eV, force convergence criterion of
0.01 eV/\AA~and strain convergence of 0.1 kbar. The DFT-optimized
crystal structures are in good agreement with the experimental
$P4/mmm$ structures ($a^{\rm{DFT}}$ = 3.91~\AA, $a^{\rm{EXP}}$ =
3.92~\AA, $c^{\rm{DFT}}$ = 3.31~\AA, $c^{\rm{EXP}}$ = 3.28~\AA). To
describe the checkerboard antiferromagnetic ordering, we expand the
cell to a $\sqrt{2}\times\sqrt{2}\times1$ supercell. To describe the
$\mathbf{q}=\left(\frac{1}{N},0,0\right)$ charge ordered state ($N$ is
an integer), we expand the cell to a $N\times 1\times 1$ supercell.

We downfold the DFT-calculated band structure of NdNiO$_2$ to a
17-orbital tight-binding model by using the MLWF
method~\cite{Marzari2012}.  The 17 orbitals include five Ni-$d$
orbitals, five Nd-$d$ orbitals, six O-$p$ orbitals and one
interstitial $s$ orbital.  This 17-orbital model can well reproduce
the non-interacting band structure of NdNiO$_2$ in an energy window of
about 15 eV around the Fermi level (see Supplementary Figure 1). A Slater-Kanamori interaction is added on the Ni-$d$
orbitals~\cite{Werner-2009}. A fully-localized-limit (FLL) double
counting correction is applied to the full interacting
model~\cite{Czyzyk1994}. We also test other double counting
correction~\cite{Haule-PRB-2014} and no qualitative change is found in the
main results. More details of the full Hamiltonian are found in Supplementary
Note 1.
We treat the two Ni $e_g$ orbitals with DMFT
method, while the filled Ni $t_{2g}$ shell is treated with a static
Hartree-Fock approximation~\cite{RN61, Karp-2020-PRX}.




When we use the DMFT method to solve the interacting 17-orbital model, we employ
the continuous-time quantum Monte Carlo (CTQMC) algorithm based on
hybridization expansion~\cite{Werner2006, Gull2011}. The impurity
solver is developed by K. Haule~\cite{Haule2007}. For each DMFT
iteration, a total of 1 billion Monte Carlo samples are collected to
converge the impurity Green function and self energy. We set the
temperature to be 116 K. We check all the main results at a lower
temperature of 58 K and no significant difference is found.
For the CO state, we allow all the Ni sites to be inequivalent and
the DMFT self-consistent condition involves the self-energies
of all inequivalent Ni atoms. 

To obtain the spectral functions, the imaginary axis self energy is
continued to the real axis by using the maximum entropy
method~\cite{Silver1990}. Then the real axis local Green function is
calculated using the Dyson equation. A $90 \times 90\times 90$
\textbf{k}-point mesh is used to converge the spectral function of the
17-orbital model.

The full charge-self-consistency does not qualitatively change the
main results of charge transfer and charge disproportionation in the
calculations~\cite{Singh-CPC-2021}. To relieve the computational burden,
the results shown in the main text are from one-shot DFT+DMFT calculations.

We also apply the above relevant parameters when using DMFT to
solve the low-energy effective model. 

\subsection*{Low-energy effective model}

\begin{table}[h]
  \caption{\label{tab1}The onsite energy and hopping matrix elements in the
    2-orbital model. All the units are eV.}
  \begin{tabular}{c|c|c|c}
\hline
\hline
in $\epsilon_d(\textbf{k})$ & $t^{100}_d = -0.370$ & $t^{110}_d = 0.060$ & $t^{200}_d = -0.020$ \\\hline
in $\epsilon_s(\textbf{k})$ & $E_{ds} = 0.70$ & $t^{100}_s = 0.018$ & $t^{110}_s = -0.049$ \\\hline
        & $t^{101}_s = -0.191 $ & $t^{111}_s = 0.075$ & $t^{001}_s = -0.237$ \\\hline
in $V_{ds}(\textbf{k})$    & $t^{100}_{ds} = -0.052$  &   &\\\hline
\hline
  \end{tabular}
\end{table}

The low-energy effective model to describe the CO phenomenon
is a 2-orbital model, which can be compactly
written as:

\begin{equation}
  \label{eqn:Xmodel1} \hat{H}  = \sum_{\textbf{k} \sigma} \hat{\Psi}^{\dagger}_{\textbf{k}\sigma}\mathcal{H}_0(\textbf{k})\hat{\Psi}_{\textbf{k}\sigma}+ U_d\sum_i \hat{n}^d_{i\uparrow}\hat{n}^d_{i\downarrow} - \hat{V}^{\rm{dc}}
\end{equation} 
where $\hat{\Psi}^{\dagger}_{\textbf{k}\sigma} = (\hat{d}^{\dagger}_{\textbf{k}\sigma}, \hat{s}^{\dagger}_{\textbf{k}\sigma})$ are the creation operators on Ni-$d_{x^2-y^2}$ and effective-$s$ orbitals with momentum $\textbf{k}$ and spin $\sigma$.
$\hat{n}^d_{i\sigma} = \hat{d}^{\dagger}_{i\sigma}\hat{d}_{i\sigma}$ is the occupancy
operator of Ni-$d_{x^2-y^2}$ orbital at site $i$ with spin $\sigma$.
$\mathcal{H}_0(\textbf{k})$ is a $2\times 2$ matrix:
\begin{equation}
  \label{eqn:Xmodel2}
\mathcal{H}_0(\textbf{k}) = \left[\begin{matrix}
\epsilon_{d}(\textbf{k}) & V_{ds}(\textbf{k})\\
V^{*}_{ds}(\textbf{k}) & \epsilon_{s}(\textbf{k})
\end{matrix}\right]
\end{equation}
The energy dispersion and hybridization terms are:
\begin{equation}
  \label{eqn:Xmodel3} \epsilon_d(\textbf{k}) = 2t^{100}_d \left(\cos k_x +\cos k_y \right) + 4t^{110}_d \cos k_x \cos k_y + 2t^{200}_d\left(\cos 2k_x + \cos 2k_y \right)
\end{equation}
\begin{eqnarray}
  \label{eqn:Xmodel4} \epsilon_s(\textbf{k}) = E_{ds} + 2t^{100}_s \left(\cos k_x+\cos k_y\right) + 2t^{001}_s\cos k_z + 4t^{110}_s \cos k_x \cos k_y 
 \\\nonumber
+ 4t^{101}_s\left(\cos k_x + \cos k_y \right) \cos k_z + 8t^{111}_s\cos k_x \cos k_y \cos k_z  
\end{eqnarray}
\begin{equation}
\label{eqn:Xmodel5} V_{ds}(\textbf{k}) = 2t^{100}_{ds}\left(\cos k_x - \cos k_y \right)(1+e^{-ik_z})
\end{equation} 
The onsite energy difference $E_{ds}$ and hopping parameters are obtained
by fitting to the near-Fermi-level DFT band structure of NdNiO$_2$ (see
Supplementary Figure 1). The
fitted values are shown in Table~\ref{tab1}. We use the FLL
double counting and $V^{\rm{dc}}$ reads:

\begin{equation}
\label{eqn:Xmodel6} V^{\rm{dc}} = U_d\left(N_d-\frac{1}{2}\right)
\end{equation}

\subsection*{Total energy}

For the low-energy effective model, we calculate the total energy
within the DFT+DMFT method. The total energy in the DFT+DMFT
calculations has the following expression ~\cite{Kotliar-RMP-2006,Hyowon2014a}:
\begin{equation}
\label{eq1} E^{\mathrm{DFT+DMFT}} = E^{\mathrm{DFT}} + E^{\mathrm{DMFT}} - E^{\mathrm{dc}}
\end{equation}
Here $E^{\rm{DFT}}$ is:
\begin{equation}
\label{eq11a} E^{\mathrm{DFT}} = E_0^{\mathrm{DFT}} - E_0^{\rm{kin}}
\end{equation}
where $E^{\rm{DFT}}_0$ is the standard DFT total energy.
$E^{\rm{kin}}_0$ is the kinetic energy of the non-interaction Hamiltonian
within the model space:
\begin{equation}
 \label{eq11b}E^{\rm{kin}}_0 = \text{Tr}\left(\hat{\mathcal{H}_0}\hat{G_0}\right) = \frac{1}{N_{\mathbf{k}}} \sum_{\mathbf{k} \sigma} \sum_{l} \epsilon_{\textbf{k}l} n_F(\epsilon_{\textbf{k}l} - \mu)
\end{equation}
where $n_F$ is the Fermi-Dirac occupancy, $l$ is the band index and $\sigma$ is the spin.

In the main text, we use the DFT optimized crystal structure for all the
states. Thus the PM, AFM and CO states all share the
same non-interacting Hamiltonian (up to a multiplication factor), we omit the
$E^{\rm{DFT}}$ in the calculation of total energy (since it is identical for
all the three states). The remaining total energy is 
\begin{equation}
\label{eq1a}E = E^{\rm{DMFT}} - E^{\rm{dc}} = E^{\mathrm{kin}} + E^{\mathrm{pot}} - E^{\mathrm{dc}}
\end{equation}
Here the DMFT kinetic energy $E^{\mathrm{kin}}$ is defined as:
\begin{equation}
\label{eq2}
E^{\text{kin}}=\text{Tr}\left(\hat{\mathcal{H}_0}\hat{G}\right) = \frac{T}{N_{\mathbf{k}}} \sum_{\mathbf{k} \sigma} \sum_{\omega_n}\sum_{m m'}
\big[\mathcal{H}_0(\textbf{k})\big]_{m m'} \big[G_{\sigma}(\textbf{k}, i\omega_n)\big]_{m' m}
\end{equation}
where $\sigma$ is the spin, $T$ is the temperature, $m, m'$ are the
indices of orbital basis. $\mathcal{H}_0(\textbf{k})$ is the
non-interacting part of the effective model Eq.~(\ref{eqn:Xmodel2})
and $G$ is the dressed lattice Green function of the interacting
model. The DMFT potential energy $E^{\textrm{pot}}$ is defined as:
\begin{equation}
  \label{eq3} E^{\text{pot}} = \frac{1}{2}\textrm{Tr}\left(\hat{\Sigma}_{\textrm{cor}}\hat{G}_{\textrm{cor}}\right)=\frac{1}{2}T\sum_{\sigma}\sum_{\omega_n}\sum_{m}e^{-i\omega_n0^{-}}
\big[\Sigma^{\sigma}_{\text{cor}}(i \omega_n)\big]_{m}\big[G^{\sigma}_{\text{cor}}(i \omega_n)\big]_{m}
\end{equation}
where $\sigma$ is the spin, $T$ is the temperature, $m$ is the index
of orbital basis. $\Sigma_{\text{cor}}$ and $G_{\text{cor}}$ are the
self-energy and the local Green function in the correlated
subspace. $E^{\mathrm{dc}}$ is the double counting energy. We
employ the FLL double counting
and $E^{\mathrm{dc}}$ reads:
\begin{equation}  \label{eq4}
E^{\mathrm{dc}} = \frac{1}{2}U_dN_d(N_d-1) 
\end{equation}
Taking derivative of $E^{\textrm{dc}}$ with respect to $N_d$ yields
the double-counting potential $V^{\textrm{dc}}$
Eq.~(\ref{eqn:Xmodel6}). In the low-energy effective model, the Hund's
term vanishes because there is only one correlated orbital
Ni-$d_{x^2-y^2}$ per site.

We note that when we study the structural distortions in the
CO state, the crystal structure changes in each calculation and so does
the DFT total energy. In that case, we need to use Eq.~(\ref{eq1}) to
calculate the DFT+DMFT total energy.

Since the total energy is calculated using CTQMC, we average the total
energy of the last ten DMFT iterations. We find that due to the highly
accurate self-energy, the standard deviation of the total energy is
about 1 meV/f.u., which is smaller than the symbol size in the
figures.


\section*{Data availability}
The data that support the findings of this study are available from
the corresponding author upon reasonable request. Source data are provided with this paper.

\section*{Code availability}
The electronic structure calculations were performed using the proprietary
code VASP~\cite{Kresse1996}, the open-source code Wannier90
~\cite{mostofi2008wannier90} and the open-source impurity solver
implemented by Kristjan Haule at Rutgers University
(http://hauleweb.rutgers.edu/tutorials/).
Both Wannier90 and Haule's impurity solver are freely
distributed on academic use under the Massachusetts Institute of Technology
(MIT) License.


\clearpage
\newpage

\begin{acknowledgments}
We thank Andrew Millis, Fu-Chun Zhang, Wei-Sheng Lee, Hyowon Park and
Pu Yu for their stimulating discussions. H.C. is supported by the
National Key R\&D Program of China (No. 2021YFE0107900) and Science
and Technology Commission of Shanghai Municipality under grant number
23ZR1445400. This work is also supported by the National Natural
Science Foundation of China (Grant No. 11774236, No. 11974397), the
Ministry of Science and Technology of China (No. 2017YFA0302902,
No. 2017YFA0303103) and the Strategic Priority Research Program of the
Chinese Academy of Sciences (Grant No.  XDB33010100). High-Performance-Computing of New York University (NYU-HPC) provides the computational resources.
\end{acknowledgments}

\section*{Author contributions}
H.C., Y.Y. and G.Z. conceived the project. H.C. and H.L. constructed the effective model and performed the calculations. H.C., Y.Y. and G.Z. wrote the manuscript. All the authors participated in the discussion.

\section*{Competing interests}
The authors declare no competing interests.

\clearpage
\newpage

\clearpage
\newpage

\clearpage
\newpage

\clearpage
\newpage

\newpage
\clearpage

\newpage
\clearpage

\end{document}